\documentclass[12pt, onecolumn, draftclsnofoot]{IEEEtran}
\usepackage[T1]{fontenc}

\ifCLASSINFOpdf

\else

\fi

\usepackage{amsmath}
\usepackage{amssymb}

\usepackage[cmintegrals]{newtxmath}

\usepackage{array}

\usepackage{fixltx2e}

\usepackage{stfloats}
\usepackage{epsfig}
\usepackage{times}
\usepackage{epsfig}
\usepackage{graphicx}
\usepackage{color}
\usepackage{algorithm}
\usepackage{algpseudocode}
\usepackage{pifont}
\usepackage{relsize}
\usepackage{verbatim}
\usepackage{csquotes}
\usepackage{lettrine}

\DeclareMathOperator{\Pro}{\mathbb{P}}

\newcommand{\te}[1]{\text{#1}}
\newcommand{\ms}[1]{\mathsf{#1}}

\newtheorem{Proposition}{\textbf{Proposition}}

\definecolor{orange}{RGB}{255, 165, 0}

\usepackage{cancel}

\def\UAVone{\mathsf{UAV_1}}
\def\UAVtwo{\mathsf{UAV_2}}
\def\Cellone{\mathsf{Cell_1}}
\def\Celltwo{\mathsf{Cell_2}}

\begin{document}

\title{Joint Configuration of Transmission Direction and Altitude in UAV-based Two-Way Communication}


\author{Wenqian~Huang, Dong Min Kim, Wenrui~Ding, and Petar Popovski
\thanks{W. Huang and W. Ding are with School of Electronic and Information Engineering, Beihang University, China
(email: \{huangwenqian, ding\}@buaa.edu.cn).}
\thanks{D. M. Kim and P. Popovski are with the Department of Electronic Systems, Aalborg University, Denmark
(email: \{dmk, petarp\}@es.aau.dk).}%
}

\maketitle

\begin{abstract}
When considering unidirectional communication for unmanned aerial vehicles (UAVs) as flying Base Stations (BSs), either
uplink or downlink, the system is limited through the co-channel interference that
takes place over line-of-sight (LoS) links. This paper considers two-way communication
and takes advantage of the fact that the interference among the ground devices takes
place through non-line-of-sight (NLoS) links. UAVs can be deployed at the high
altitudes to have larger coverage, while the two-way communication allows to configure
the transmission direction.
Using these two levers, we show how the system throughput can be maximized for a given
deployment of the ground devices.
\end{abstract}


\section{Introduction}

A number of recent works have considered the use of
 Unmanned aerial vehicles (UAVs) as flying Base Stations (BSs) to provide data services to ground users, see~\cite{mozaffari2018tutorial} and the references therein. Compared to a traditional terrestrial BS, UAV can provide higher capacity since line-of-sight (LoS) wireless communication link can be easily established in UAV-ground channel~\cite{al2014modeling}. Moreover, multiple UAVs are considered to enhance coverage as well as throughput with the increasing demand of densely deployment for UAVs~\cite{mozaffari2016efficient}.
Despite the benefits, there are also a number of challenges, such as cell partition according to user distribution, UAVs' altitudes adjustment and interference management, etc. Furthermore, while the LoS link UAV-ground is beneficial for the useful signal, it should be noted that, when one-way communication (either uplink or downlink) is considered, the co-channel interference always comes through the LoS link as well.

Authors in \cite{Kosmerl2014base} used evolutionary algorithms in order to find the
optimal placement of UAVs to help terrestrial stations to provide wireless services in
disaster relief scenarios. Results in \cite{Kosmerl2014base} shows that increasing the
number of UAVs will inevitably lead to interference due to overlapping areas. In
\cite{mozaffari2016efficient}, in order to maximize the downlink coverage, the authors
proposed an efficient deployment method, using circle packing theory, for multiple UAVs
while ensuring that the coverage areas of UAVs do not overlap. Authors in
\cite{lyu2017blocking} considered uplink transmission from UAVs to ground BSs. A
communication block is defined for a UAV if it cannot find a BS or there are
other BSs stations in its main lobe serving other UAVs, and hence suffer from
strong co-channel interference. This blocking probability is minimized by adjusting the
UAV's altitude and/or beamwidth.  Hence, in general, UAVs with the directional antennas
can have a larger coverage to serve more ground users at the relatively higher altitudes
but at the cost of more path loss and interference.
The works described above are focused on one-way communication scenario, either uplink or downlink. The method that is commonly used to mitigate co-channel interference is to avoid overlapping areas through altitude control~\cite{mozaffari2016efficient} or joint control of the directional antenna's beamwidth as well as altitude~\cite{lyu2017blocking}.


In this paper, we consider a two-way communication scenario and mitigate the co-channel interference by using the transmission direction as an additional degree of freedom. Our study reveals that adapting the  transmission directions for multiple UAV-user links can reduce the co-channel interference, which means that the UAVs can be deployed at relatively high altitudes, thereby enlarging their coverage.  Furthermore, the altitudes can also be adjusted according to the topology of ground users, thus improving the system-level throughput.

\begin{figure}[!tp]
	\begin{center}
		\vspace{-0.1cm}
		\includegraphics[width=9cm]{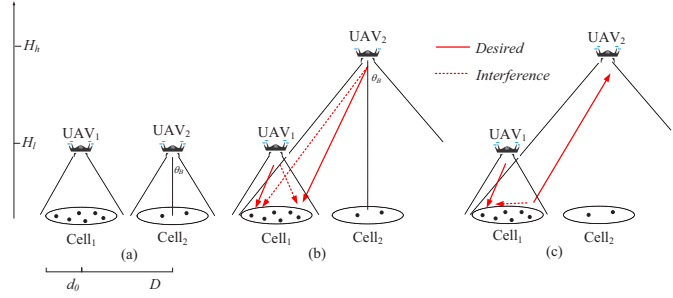}
		\vspace{-0.3cm}
		\caption{Asymmetrical scenario with (a) low-altitude deployment of UAVs; (b) different-altitude deployment of UAVs using the same transmission direction; (c) different-altitude deployment of UAVs using different transmission directions }\vspace{-.32cm}
		\label{high}
	\end{center}\vspace{-.12cm}
\end{figure}

\begin{figure}[!tp]
	\begin{center}
		\vspace{-0.1cm}
		\includegraphics[width=7cm]{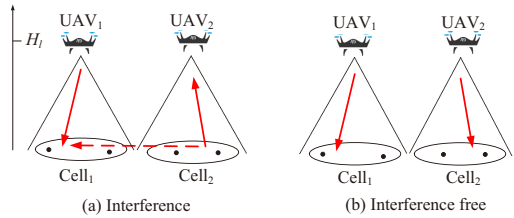}
		\vspace{-0.3cm}
		\caption{Symmetrical scenario with low-altitude deployment of UAVs using (a) different transmission directions; (b) the same transmission direction.}\vspace{-.32cm}
		\label{equal}
	\end{center}\vspace{-.12cm}
\end{figure}

Consider deployment of two UAVs to serve two cells, see Fig.~\ref{high}. Note that, here we
define a \emph{cell} to be a spatial region of a fixed size in which the users are deployed
randomly, rather than the region that is under a coverage of a specific UAV. The two
cells on Fig.~\ref{high}(a) are asymmetric in a sense that the device/user population in the
$\Cellone$ is much larger compared to the one in $\Celltwo$. Due to this, in the
low-altitude deployment from Fig.~\ref{high}(a), the communication resources of $\UAVtwo$
will be poorly used if each of the users in $\Celltwo$ is only intermittently active.
This can be amended by placing $\UAVtwo$ at a higher altitude, as in Fig.~\ref{high}(b).
However, if a device communicates in the downlink with $\UAVone$ while another device
communicates with $\UAVtwo$, then a co-channel interference occurs, as shown on
Fig.~\ref{high}(b). The same happens when both UAVs transmit in the uplink to 
different devices. However, if the two active links operate in
opposite directions, as shown in Fig.~\ref{high}(c), then the co-channel
interference can be mitigated, as it occurs over a ground-to-ground NLoS link.

If the user population in the two cells is more symmetric, as in Fig.~\ref{equal}, the two UAVs
do not need to cooperate: each UAV should exclusively serve one cell while being deployed at a lower altitude. This minimizes the path loss, while avoiding interference. In this scenario, transmitting in the same, rather than opposite, direction in both cells is better.  This is because ground users will still suffer interference from other ground users using the different directions while there will be an interference-free scenario for all nodes using the same transmission direction~\cite{mozaffari2016efficient}.

The adjustment of the transmission direction has been introduced in \cite{popovski2015interference} under the
name of interference spin. This concept is here enriched by adjusting the UAV altitude, resulting in improvements in system throughput.

\section{System Model}

We consider dual UAVs, as aerial BSs, are deployed to serve two circle cells,
$\Cellone$ and $\Celltwo$, with the same radius $d_0$ and the distance between
centers of two cells is denoted as $D$.

A set $\mathcal{N}=\{1,2...N\}$ of $N$ users are assumed to be uniformly distributed in
each cell with Poisson probabilistic activation during a transmission frame. That is, $\Cellone$ and $\Celltwo$
respectively include $K_1$=Pois($\lambda_1$) $\in \mathcal{N}$ and $K_2$=Pois($\lambda_2$) $\in \mathcal{N}$ active ground users, where $\lambda_1 \in \mathcal{N}$,
$\lambda_2 \in \mathcal{N}$ denote the densities of active user in two
cells.\footnote{The actual distribution is binomial distribution. We approximate it using
Poisson distribution as the number of users is sufficiently large and the active
probability for each user is sufficiently small.} A new transmission frame starts when the whole set active users have already been served. In the two-dimensional (2D) horizontal
plane, we deploy two UAVs at the centers of two cells, respectively. The
three-dimensional (3D) coordinates of $\UAVone$ and $\UAVtwo$ can be denoted
by, respectively, $\boldsymbol{s}_{1}=(x_{0},y_0, h_1)$ and $\boldsymbol{s}_{2}=(x_{0}',y_{0}', h_2)$.
The coordinate of ground user $G_{j}, j \in \mathcal{K}=\{1, 2,..., 2\te{N}\}$ can be
denoted as $\boldsymbol{w}_{j}=(x_{j},y_j,0)$. Transmit powers of UAV and ground users
are denoted as $P_u$ and $P_g$, respectively.

All devices operate in half-duplex mode. Both UAVs use the same frequency band, normalized to 1 Hz and each UAV-user operates in time division duplex (TDD) mode. Load-balanced two-way communication takes place in two successive time slots: the first slot is used in one direction, either downlink or uplink, and the successive slot in the opposite direction.

At the start of each transmission frame, the topology of active ground users can be known by a cloud center, e.g., a macro BS, and then the optimal transmission parameters can be calculated and sent to UAVs as well as ground users~\cite{dowhuszko2013decentralized}. Compared to data transmission time, the parameter calculation time can be ignored in each transmission frame.
The whole system consists of two two-way links: link $L_1$ with $\UAVone$ and link $L_2$ with $\UAVtwo$. All links are slot-synchronous.

We consider the directional antenna for each UAV and the antenna gain of $\UAVone$, for
example, seen by $G_j$ is approximately by~\cite{mozaffari2016efficient}:
\begin{IEEEeqnarray}{rCl}\label{antenna gain}
g (d_{1j}) & = & \left\{
	\begin{array}{lcl}
	\frac{g_0}{\Phi^2_{B}}, &d_{1j}= \parallel \boldsymbol{s}_{1}-\boldsymbol{w}_{j} \parallel \le h_1/\mathrm{cos}\Phi_B; \\
	0,       					& \mbox{otherwise}.
	\end{array}
\right.
\end{IEEEeqnarray}
where $g_0=\frac{30000}{2^2} \times (\frac{\pi}{180})^2 \approx 2.2846$ and $\Phi_B \in
(0,\frac{\pi}{2})$ denotes antenna's half beamwidth
\cite{lyu2017blocking}. The term $d_{1j}$ denotes the distance between
$\UAVone$ and ground user $j$ in $\Cellone$. The omnidirectional antenna is assumed for
each ground user with unit gain $g_0$.

The UAV-ground channel model is considered as LoS link while the ground-ground channel model is a NLoS link due to obstacles on the ground. For ease of illustrating our work, large-scale fading with path loss and shadow fading is considered for two kinds of channels, and the fading channels are assumed to be constant in each time slots but vary between different slots. This is sufficient to support our concept, but the work can be extended to other fading models as well.  
The received signal power of $G_j$ from $\UAVone$ is given by~\cite{al2014modeling}:
\begin{IEEEeqnarray}{rCl}\label{Pij}
P_{1j}(d_{1j})=\frac{P_u g(d_{1j}) }{\psi_{\mathrm{LoS}}}(\frac{4\pi f_c}{c} d_{1j})^{-n_{\mathrm{LoS}}}
\end{IEEEeqnarray}
where $n_{\mathrm{LoS}}=2$ is path loss exponent for LoS propagation, $f_c$ is the carrier frequency and $c$ is the speed of light.
$\psi_{\mathrm{LoS}} \sim N(\mu_{\mathrm{LoS}},\sigma^2_{\mathrm{LoS}})$ is shadow fading with normal distribution for LoS link.

Similar to (\ref{Pij}), assuming channel reciprocity, the received signal powers of
$\UAVone$ and the ground user $G_{j'}$ from $G_j$ can be respectively given by
\begin{IEEEeqnarray}{rCl}
P_{j1}(d_{j1})& =& \frac{P_g g_0}{\psi_{\mathrm{LoS}}}(\frac{4\pi f_c}{c}d_{j1})^{-n_{\mathrm{LoS}}}\label{Pj1}\\
P_{jj'}(d_{jj'})& =& \frac{P_g g_0 }{\psi_{\mathrm{NLoS}}}(\frac{4\pi f_c}{c} d_{jj'})^{-n_{\mathrm{NLoS}}}\label{Pjj'}
\end{IEEEeqnarray}
where $j' \in \mathcal{K}$ and $j' \ne j$, $n_{\mathrm{NLoS}}=4$ is path loss exponent for NLoS propagation. $\psi_{\mathrm{NLoS}} \sim N(\mu_{\mathrm{NLoS}},\sigma^2_{\mathrm{NLoS}})$ is shadow fading with normal distribution for NLoS link.

In this paper, we consider that each UAV can be deployed at two specific altitudes: either a low altitude $H_l=d_0/\text{tan}(\Phi_B)+h_0$ or a high altitude $H_h=(d_0+D)/\text{tan}(\Phi_B)$, where $h_0 \ge 0$ is set to make sure that the UAV with low altitude will not fall into antenna main lobe of the other UAV with the high altitude, thus the interference between UAVs can be ignored.

\section{ System Throughput and Optimal Configuration}\label{throughput}

\subsection{Two-way sum-rate}

The \emph{interference spin} or, for short, \emph{spin} of link $L_1$ can be defined
as~\cite{popovski2015interference}: $p_{1}=0$ if downlink takes place in the odd slot and
uplink takes place in the even slot; $p_{1}=1$ vice versa. Furthermore, the relative spin, to illustrate the
interference between two-way links $L_1$ and $L_2$, can be defined as $r=p_1 \oplus p_2$, where $\oplus$ is an XOR operator. In summary, $r$=1 means two links use different
communication directions while $r$=0 means they use the same direction.

With this, the received signal-to-interference-and-noise-ratio (SINR) for any link can be
expressed as the function of $r$. We consider two co-channel links: $L_1$ with $\UAVone$ and $G_j$; $L_2$ with $\UAVtwo$ and $G_{j'}$. According to geometry, we have $h_1 \leqslant d_{1j} \leqslant h_1/\mathrm{cos}\Phi_B$, $h_2  \leqslant  d_{2j} \leqslant h_2/\mathrm{cos}\Phi_B$.

Consider the the co-channel users $G_j$ and $G_{j'}$ that are located in different cells as shown in Fig.~\ref{equal}. The received signal in link $L_1$ at $G_j$ from $\UAVone$ experiences co-channel interference from $\UAVtwo$ only when $\UAVtwo$ is deployed at a high altitude and the relative spin is $r=0$.  
On the other hand, interference from the co-channel user $G_{j'}$ is present when $r=1$. Thus, in this scenario, the received SINR of downlink at $G_j$ from $\UAVone$ is given as
\begin{IEEEeqnarray}{lCl}
\mathsf{SINR}^{1}_{ug,d}&=&\frac{P_{1j}(d_{1j})}{r P_{j'j}(d_{jj'})+t(h_2)(1-r)P_{2j}(d_{2j})+ \sigma^2} \label{SINRij}\nonumber\\
&\geqslant &  \frac{P_{1j}(h_1/\mathrm{cos}\Phi_B)}{r P_{j'j}(d_{min})+t(h_2)(1-r)P_{2j}(h_2)+ \sigma^2}
\triangleq  \overline{\mathsf{SINR}}^{1}_{ug}(r, h_1,h_2)\label{SINRij apprx define}
\end{IEEEeqnarray}
where  $\sigma^2$ denotes the power of additive white Gaussian (AWGN) and $d_{min}=2d_0/N$ is approximated as the minimal distance between ground users by simply assuming that users are deployed with equidistant intervals. The binary indicator
$t=1$ means UAV is deployed at the $H_h$ while $t=0$ means UAV is deployed at the $H_l$,
and it can be expressed as
\begin{IEEEeqnarray}{rCl}\label{altitudes parameter}
	t(h)& = &  \frac{h-H_l}{H_h-H_l}=
	\left\{
	\begin{array}{lcl}
		1, & h=H_h; \\
		0,       					& h=H_l.
	\end{array}
	\right.
\end{IEEEeqnarray}

For simplicity and similar to (\ref{SINRij apprx define}), in the following text we adopt the lower bounds on SINRs~\cite{lyu2017blocking}.
For the received signal of uplink at $\UAVone$  from $G_{j}$, it suffers from interference from the co-channel user $G_{j'}$ only when $\UAVone$ is deployed at the high altitude and $r=1$. The received SINR at $\UAVone$ is:
\begin{IEEEeqnarray}{lCl}
\mathsf{SINR}^{1}_{gu,d}&=&\frac{P_{j1}(d_{j1})}{t(h_1)(1-r)P_{j'1}d_{j'1}+ \sigma^2} \label{SINRji}
 \geqslant  \frac{P_{j1}(h_1/\mathrm{cos}\Phi_B)}{t(h_1)(1-r)P_{j'1}(h_1)+ \sigma^2} \triangleq   \overline{\mathsf{SINR}}^{1}_{gu}(r, h_1)\label{SINRji apprx define}
 \IEEEeqnarraynumspace
\end{IEEEeqnarray}

When the co-channel users are located in the same cells as shown in Fig.~\ref{high}, the received signal in downlink at $G_j$ from $\UAVone$ always gets co-channel interference from $\UAVtwo$ when the two links use the same direction $r=0$. Furthermore, it still gets interference from the co-channel user $G_{j'}$ when $r=1$. It is analogous for the received signal in uplink. Then, the lower bounds on the SINR in this scenario are given, respectively, by:
\begin{IEEEeqnarray}{lCl}
\overline{\mathsf{SINR}}^{1}_{ug,s}(r, h_1,h_2)	=  \frac{P_{1j}(h_1/\mathrm{cos}\Phi_B)}{rP_{j'j}(d_{min})+(1-r)P_{2j}(h_2)+ \sigma^2}
 \IEEEeqnarraynumspace\label{SINRijs apprx define}\\
\overline{\mathsf{SINR}}^{1}_{gu,s}(r,h_1)= \frac{P_{j1}(h_1/\mathrm{cos}\Phi_B)}{(1-r)P_{j'1}(h_1)+ \sigma^2}
 \IEEEeqnarraynumspace\label{SINRjis apprx define}
\end{IEEEeqnarray}

The SINRs for link $L_2$:  $\ms{\overline{SINR}}^{2}_{ug,d}$, $\ms{\overline{SINR}}^{2}_{gu,d}$, $\ms{\overline{SINR}}^{2}_{ug,s}$, $\ms{\overline{SINR}}^{2}_{gu,s}$ can be derived in a similar way. Then, the maximum achievable \textit{co-channel two-way sum-rate} for co-channel links $L_1$ and $L_2$ in two scenarios can be expressed, respectively, as:
\begin{IEEEeqnarray}{lCl}
	R_{d}^{(c)}&(r,h_1, h_2)	
	=\text{log}_2 \left(
	1+\overline{\ms{SINR}}^{1}_{ug,d} \right)+\text{log}_2 \left(
	1+\overline{\ms{SINR}}^{1}_{gu,d}
	\right)	\nonumber\\
	&+ \text{log}_2 \left(
	1+\overline{\ms{SINR}}^{2}_{ug,d} \right)	+\text{log}_2 \left(
	1+\overline{\ms{SINR}}^{2}_{gu,d}
	\right) \label{Sum rate ijd}\\
	R_{s}^{(c)}&(r,h_1, h_2)
	=\text{log}_2 \left(
	1+\overline{\ms{SINR}}^{1}_{ug,s} \right)+\text{log}_2 \left(
	1+\overline{\ms{SINR}}^{1}_{gu,s}
	\right)\nonumber\\
	&+  \text{log}_2 \left(
	1+\overline{\ms{SINR}}^{2}_{ug,s} \right)+\text{log}_2 \left(
	1+\overline{\ms{SINR}}^{2}_{gu,s}
	\right) \label{Sum rate ijs}
\end{IEEEeqnarray}

The users, without co-channel link, can be individually served without co-channel interference, and then the \textit{individual two-way sum-rate}, in link $L_1$, can be expressed as
\begin{IEEEeqnarray}{lCl} \label{Sum rate Ri1}
R^{(i)}(h_1)=\text{log}_2 \left(
1+\overline{\ms{SNR}}^{1}_{ug}(h_1)\right)+\text{log}_2 \left(
1+\overline{\ms{SNR}}^{1}_{gu}(h_1)
\right)
 \IEEEeqnarraynumspace
\end{IEEEeqnarray}
where $\overline{\ms{SNR}}^{1}_{ug}$ and $\overline{\ms{SNR}}^{1}_{gu}$ can be respectively expressed as
$\overline{\mathsf{SNR}}^{1}_{ug}(h_1)=
\frac{P_{1j}(h_1/\mathrm{cos}\Phi_B)}{\sigma^2}$ and
$\overline{\mathsf{SNR}}^{1}_{gu}(h_1)=
\frac{P_{j1}(h_1/\mathrm{cos}\Phi_B)}{ \sigma^2}
 \label{SNRji apprx defineu}$.

The individual two-way sum-rate $R^{(i)}(h_2)$ in link $L_2$ can be expressed in a similar way.

\subsection{Transmission scheme and average throughput}\label{scheme}

Transmission scheme for this UAV-based two-way system will divided into three steps: 1) $\UAVone$ and $\UAVtwo$ serve users in their corresponding cells until there are no co-channel users in the cells; 2) If two UAVs are both deployed at the low altitudes, e.g., Fig.~\ref{high}(a), the remaining users in $\Cellone$ are individually served by $\UAVone$; if one of UAVs is deployed at the high altitude, e.g., $\UAVtwo$ in Fig.~\ref{high}(b), it can help $\UAVone$ to serve the remaining users in $\Cellone$; 3) If there is still individual user in $\Cellone$ after cooperatively served by $\UAVtwo$ in Fig.~\ref{high}(b), then it will be individually served by its corresponding UAV ($\UAVone$).
Thus the number of pairs of co-channel users and individual users vary with the UAV altitudes.
As explained earlier, three cases of potential optimal UAV altitudes can reach the maximal throughput with different active users in two cells.

Defining $k=K_1-K_2$, number of pairs of co-channel users in different cells $a_d$, number of pairs of co-channel users in the same cell $a_s$ and number of individual served users $b$ can be expressed as
\begin{IEEEeqnarray}{lCl}\label{ad+}
a_{d}^{(+)}(h_1,h_2|k>0)
=K_2\\
a_{d}^{(-)}(h_1,h_2|k \le 0)
= K_2+k\\
	a_{s}^{(+)}(h_1,h_2|k>0)
	\!=\!\! \left\{
	\begin{array}{lcl}
		\!\!0,  h_1\!=\!H_l,h_2\!=\!H_l \mathrm{\ or\ } h_1\!=\!H_h,h_2\!=\!H_l; \\
		\!\!k, h_1=H_l,h_2=H_h.
	\end{array}
	\right.\\
	a_{s}^{(-)}(h_1,h_2|k \le 0)
	\!=\! \left\{
	\begin{array}{lcl}
		\!\!0,  h_1\!=\!H_l,h_2\!=\!H_l \mathrm{\ or\ } h_1\!=\!H_l,h_2\!=\!H_h; \\
		\!\!-k, h_1=H_h,h_2=H_l.
	\end{array}
	\right.\\
b^{(+)}(h_1,h_2|k>0)
\!=\!\! \left\{
	\begin{array}{lcl}
	\!\!k,  h_1\!=\!H_l,h_2\!=\!H_l \mathrm{\ or\ } h_1\!=\!H_h,h_2\!=\!H_l; \\
	\!\!k-2\lfloor \frac{k}{2}\rfloor,	h_1=H_l,h_2=H_h.
	\end{array}
\right.\\
b^{(-)}(h_1,h_2|k \le 0)
\!=\!\! \left\{
	\begin{array}{lcl}
	\!\!-k,  h_1\!=\!H_l,h_2\!=\!H_l \mathrm{\ or\ } h_1\!=\!H_l,h_2\!=\!H_h; \\
	\!\!-k-2\lfloor \frac{-k}{2}\rfloor, h_1=H_h,h_2=H_l.
	\end{array}
\right.
\end{IEEEeqnarray}


Then, considering Poisson probabilistic activation of ground users, the average
throughput of this system can be given by the following proposition.

\begin{Proposition}\label{Proposition1}
The average throughput of UAV-based two-way communication system can be expressed as
\begin{IEEEeqnarray}{lCl}\label{Average throughput}
C (r, h_1, h_2, \lambda_1, \lambda_2) & =& \sum_{k = 1}^{N} \text{P}(k,\lambda_1, \lambda_2) \frac{\mathlarger{\sum}_{K_2+k,K_2 \in \mathcal{N}}{N \choose K_2+k}{N \choose K_2} c^{(+)}(r, h_1, h_2|k,K_2)}{\mathlarger{\sum}_{K_2+k,K_2 \in \mathcal{N}}{N \choose K_2+k}{N \choose K_2}}\nonumber\\
&+& \sum_{k = -N}^{0} \!\text{P}(k,\lambda_1, \lambda_2)\frac{\mathlarger{\sum}_{K_2+k,K_2 \in \mathcal{N}}\!\!{N \choose K_2+k}{N \choose K_2} c^{(-)}(r, h_1, h_2|k,K_2)}{\mathlarger{\sum}_{K_2+k,K_2 \in \mathcal{N}}{N \choose K_2+k}{N \choose K_2}}
\end{IEEEeqnarray}
where $r \in \{0,1\}$ and $h_1, h_2 \in \{H_h, H_l\}$. $\text{P}(k,\lambda_1, \lambda_2)$,  $c^{(+)}(\cdot|k)$ and $c^{(-)}(\cdot|k)$ are respectively expressed as
\begin{IEEEeqnarray}{rll}
\text{P}(k, \lambda_1, \lambda_2)=e^{-(\lambda_1+\lambda_2)}(\frac{\lambda_1}{\lambda_2})^{k/2} I_{k}(2\sqrt{\lambda_1 \lambda_2})\label{P1 Poisson}\\
c^{(+)}(r,h_1,h_2|k, K_2)=\frac{a_{d}^{(+)}R_{d}^{(c)}+a_{s}^{(+)}R_{s}^{(c)}+b^{(+)}R^{(i)}}{2(a_{d}^{(+)}+a_{s}^{(+)}+b^{(+)})}\label{Average throughput c+}\\
c^{(-)}(r,h_1,h_2|k, K_2)=\frac{a_{d}^{(-)}R_{d}^{(c)}+a_{s}^{(-)}R_{s}^{(c)}+b^{(-)}R^{(i)}}{2(a_{d}^{(-)}+a_{s}^{(-)}+b^{(-)})}\label{Average throughput c-}
\end{IEEEeqnarray}
where $I_{k}(z)$ is the modified Bessel function of the first kind.
\end{Proposition}

\begin{IEEEproof}
$\text{P}(k,\lambda_1, \lambda_2)=\Pro\{K_1-K_2=k\}$ is defined as the probability that
$K_1$ and $K_2$ users, with difference $k \in [-N,N]$ users, are respectively active in
$\Cellone$ and $\Celltwo$. Given the Poisson distributed population of active users,
$\Pro\{K_1-K_2=k\}$, following the Skellam distribution~\cite{skellam1946frequency}, can
be given by \eqref{P1 Poisson}.

Given the difference active users $k$, the whole active cases should be $\sum_{K_2+k,K_2 \in \mathcal{N}}{N \choose K_2+k}{N \choose K_2}$.
For each case, that is with given $K_2$ as well as $k$, the average throughput $c^{(+)}(r, h_1, h_2|k,K_2)$ can be calculated by three parts: sum-rates of the co-channel users in different cells, the co-channel users in the same cell and the individual users. As shown in (\ref{Average throughput c+}), the numerator denotes the the total two-way sum-rate for all users. The denominator is the number of time slots to serve the whole users.
The derivation for $c^{(-)}(\cdot)$ is the same.
\end{IEEEproof}

\subsection{Spin and altitude configuration}

It can be seen from (\ref{Average throughput}) that the average throughput depends on three key parameters: spin, UAV altitudes and user densities.
Defining parameter set $\mathcal{\eta}=\{r,h_1,h_2\}$, as described earlier, we can derive that only three configurations influence the maximal throughput according to different $\lambda_1$ and $\lambda_2$:
 $\mathcal{\eta}_1=\{1, H_l, H_h\}$, $\mathcal{\eta}_2=\{1, H_h, H_l\}$ and $\mathcal{\eta}_3=\{0, H_l, H_l\}$.

Then, the optimal configuration can be calculated by
\begin{IEEEeqnarray}{Cll}\label{simplified optimization problem}
\eta^*={\arg\max}_{\eta \in \{\eta_1,\eta_2,\eta_3\}} C(\lambda_1, \lambda_2,\eta )
\end{IEEEeqnarray}

\section{Numerical Results and Conclusions}

\begin{figure}[t]
	\begin{center}
		\vspace{-0.1cm}
		\includegraphics[width=9cm,height=6cm]{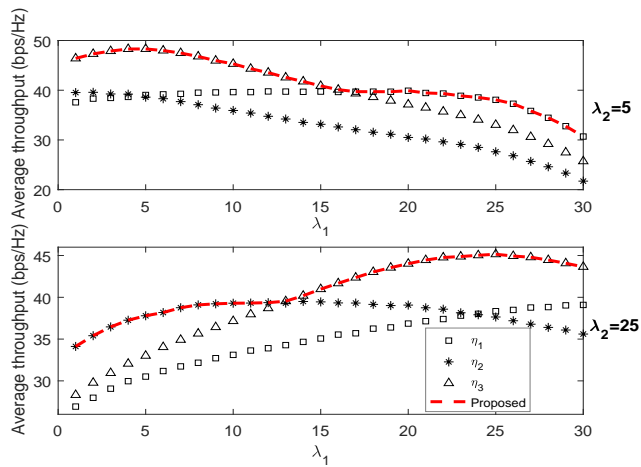}
		\vspace{-0.3cm}
		\caption{Average throughput comparison between fixed and proposed optimal configurations.}\vspace{-.32cm}
		\label{proposed}
	\end{center}\vspace{-.12cm}
\end{figure}

In the simulations, we consider the two-way communication over $f_c=2$ GHz carrier frequency with equally transmit powers for all devices ($P_u=P_g=35$ dBm) in an urban environment ($\mu_{\mathrm{LoS}}=1$ dB, $\sigma_{\mathrm{LoS}}=1$ dB, $\mu_{\mathrm{NLoS}}=30$ dB, $\sigma_{\mathrm{NLoS}}=8$ dB~\cite{al2014modeling, mozaffari2016efficient}).
Other parameters can be set as: $\sigma^2=-120$ dBm, $h_0=1$ m, $d_0=100$ m, $D=300$ m, $N=30$, $\Phi_B=\pi /3$.
Fig.~\ref{proposed} compares the throughput performance for fixed and proposed optimal configurations of spin and altitudes. Result shows that deploying two UAVs at the low altitudes and using the same direction can reach the highest throughput when two cells have relatively similar user densities. On the other hand, when the user densities in two cells have relatively large difference, the maximal throughput can be achieved by using different directions and deploying them at different altitudes, the one with less actively users in its corresponding cell at high altitude and the other at low altitude.
Moreover, Fig.~\ref{proposed} shows the optimal configuration in (\ref{simplified optimization problem}) can maximize throughput across the whole $\lambda_1$ with different $\lambda_2$.

The main conclusion is that using different transmission directions for UAV-based two-way communication can reduce the co-channel interference as much as possible and UAVs can be deployed at the high altitudes to have a larger coverage. Then, the transmission directions and UAVs' altitudes can be configured according to a given deployment of the ground users in order to maximize system throughput.



\bibliographystyle{IEEEtran}
\bibliography{Huangbibfile}
%

\end{document}